\documentclass[twocolumn,aps,prl,superscriptaddress,floatfix, nofootinbib]{revtex4-1}
\usepackage{graphicx}
\usepackage{amsmath}
\usepackage{natbib}
\usepackage{gensymb}
\usepackage{afterpage}

\begin{document}

\title{Gigahertz frequency antiferromagnetic resonance and strong magnon-magnon coupling in the layered crystal CrCl$_3$}
\author{David MacNeill*}
\affiliation{Department of Physics, Massachusetts Institute of Technology, Cambridge, MA 02139, USA}
\thanks{D.\ MacNeill and J.\ T.\ Hou contributed equally to this work.}
\author{Justin T.\ Hou*}
\affiliation{Department of Electrical Engineering and Computer Science, Massachusetts Institute of Technology, Cambridge, MA 02139, USA}
\thanks{D.\ MacNeill and J.\ T.\ Hou contributed equally to this work.}
\author{Dahlia R. Klein}
\affiliation{Department of Physics, Massachusetts Institute of Technology, Cambridge, MA 02139, USA}
\author{Pengxiang Zhang}
\affiliation{Department of Electrical Engineering and Computer Science, Massachusetts Institute of Technology, Cambridge, MA 02139, USA}
\author{Pablo Jarillo-Herrero}
\affiliation{Department of Physics, Massachusetts Institute of Technology, Cambridge, MA 02139, USA}
\author{Luqiao Liu}
\affiliation{Department of Electrical Engineering and Computer Science, Massachusetts Institute of Technology, Cambridge, MA 02139, USA}

\date{{\small \today}}

\begin{abstract}
We report broadband microwave absorption spectroscopy of the layered antiferromagnet CrCl$_3$. We observe a rich structure of resonances arising from quasi-two-dimensional antiferromagnetic dynamics. Due to the weak interlayer magnetic coupling in this material, we are able to observe both optical and acoustic branches of antiferromagnetic resonance in the GHz frequency range and a symmetry-protected crossing between them. By breaking rotational symmetry, we further show that strong magnon-magnon coupling with large tunable gaps can be induced between the two resonant modes. 
\end{abstract}
\maketitle

Antiferromagnetic spintronics is an emerging field with the potential to realize high speed logic and memory devices \cite{Jungwirth2016,Wadley2016, Bhattacharjee2018, Baltz2018, Duine2018,Olejnik2018}. Compared to ferromagnetic materials, antiferromagnetic dynamics are less well-understood \cite{cheng_spin_2014,liu_mode-dependent_2017, cheng_terahertz_2016,kamra_spin_2017}, partly due to their high instrinsic frequencies that require specialized terahertz techniques to probe \cite{bhattacharjee_neel_2018,kampfrath_coherent_2011,baierl_terahertz-driven_2016}. Therefore, antiferromagnetic materials with lower and tunable resonant frequencies are desired to enable a wide range of fundamental and applied research \cite{Duine2018}. Here, we introduce the layered antiferromagnetic insulator CrCl$_3$ as a tunable platform for studying antiferromagnetic dynamics. Due to the weak interlayer coupling of CrCl$_3$, the antiferromagnetic resonance (AFMR) frequencies are within the range of typical microwave electronics (\textless 20 GHz). This allows us to excite different modes of AFMR and to induce a symmetry-protected mode crossing with an external magnetic field. We further show that a tunable coupling between the optical and acoustic magnon modes can be realized by breaking rotational symmetry. Recently, strong magnon-magnon coupling between two adjacent magnetic layers has been achieved \cite{chen_strong_2018, klingler_spin-torque_2018}, with potential applications in hybrid quantum systems \cite{zhang_strongly_2014, bai_spin_2015,tabuchi_coherent_2015}. Our results demonstrate strong magnon-magnon coupling within a single material and therefore provide a versatile system for microwave control of antiferromagnetic dynamics. Furthermore, CrCl$_3$ crystals can be exfoliated down to the monolayer limit \cite{McGuire22017} allowing facile device integration for antiferromagnetic spintronics. 

\begin{figure}[!tbp]
\includegraphics[width=8.5 cm]{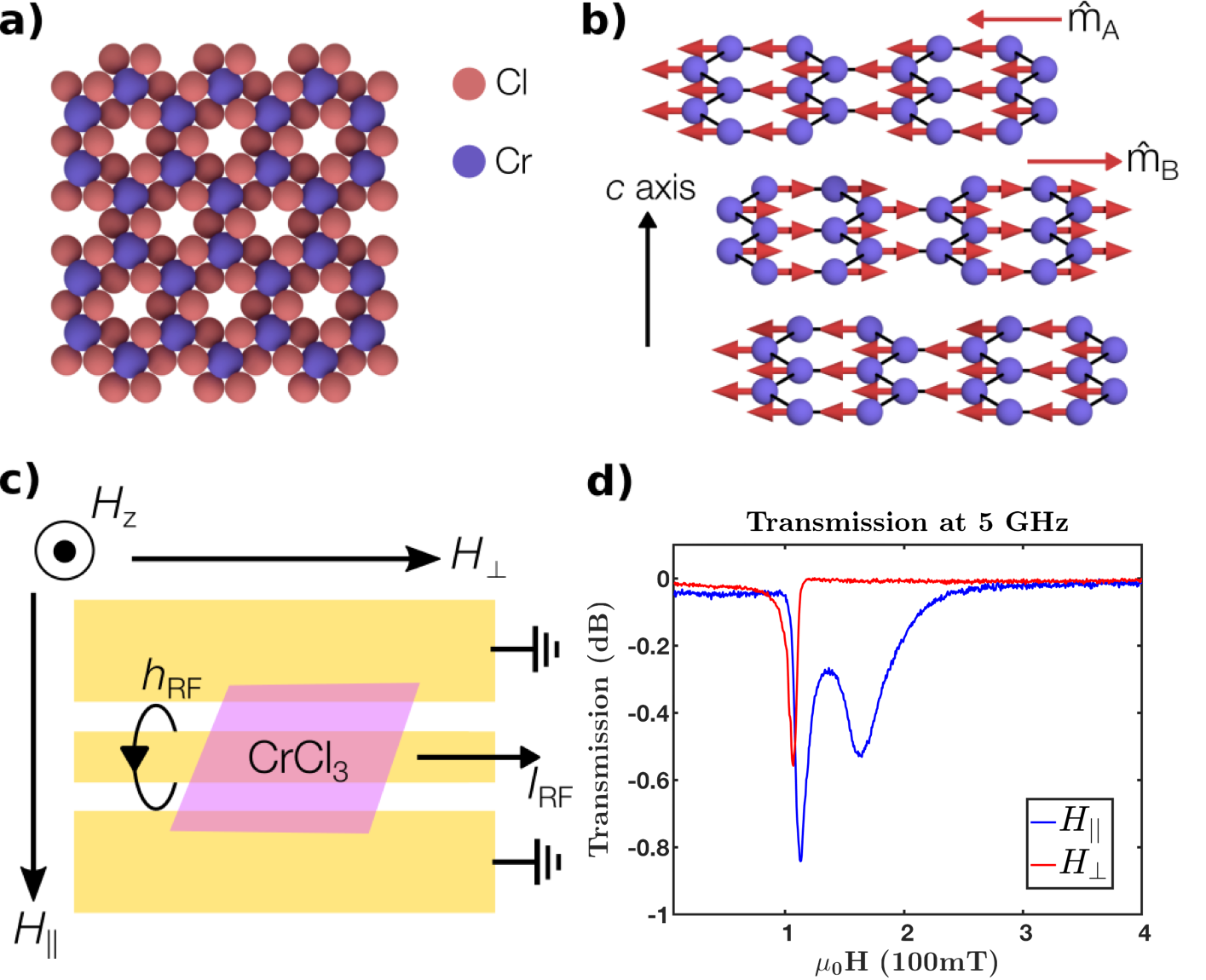}
    \caption{ {\bf a)} Crystal structure of a single CrCl$_3$ layer. Red and purple spheres represent chlorine and chromium atoms respectively. {\bf b)} Magnetic structure of bulk CrCl$_3$ below the Ne{\'e}l temperature, and without an applied magnetic field. The blue spheres represent the Cr atoms. The red arrows represent the magnetic moment of each Cr atom with parallel intralayer alignment and antiparallel interlayer alignment. The net magnetization direction alternates between layers, having direction $\hat{m}_{\mathrm{A}}$ ($\hat{m}_{\mathrm{B}}$) on layers in the A (B) magnetic sublattice. {\bf c)} Schematic of the experimental geometry featuring a coplanar waveguide (CPW) with a CrCl$_3$ crystal placed over the signal line. $H_{||}$, $H_{\perp}$, and $H_z$ are the components of the applied DC magnetic field. The microwave transmission coefficient was measured as a function of applied magnetic field and temperature. {\bf d)} Typical microwave transmission at 5 GHz and 1.56 K as a function of magnetic field applied parallel (blue) or perpendicular (red) to the in-plane RF field, showing resonances due to AFMR. The two traces were taken from different CrCl$_3$ crystals. }
    \label{fig1}
\end{figure}

The crystal and magnetic structures of CrCl$_3$ are shown in Fig. \ref{fig1}a and \ref{fig1}b \cite{McGuire2017,Gossard1961,Narath1963, Davis1964,Narath1965,Narath21965, Samuelsen1971}. Spins within each layer have a ferromagnetic nearest-neighbor coupling of about 0.5 meV, whereas spins in adjacent layers have a weak antiferromagnetic coupling of about 1.6 $\mu$eV \cite{Narath21965}. Therefore, we can consider each layer as a two-dimensional ferromagnet coupled to the adjacent layers by an interlayer exchange field of roughly 0.1 T \cite{Narath21965}. The very weak interlayer coupling implies that the field and frequency required to manipulate the antiferromagnetic order parameter (N\'{e}el vector) are orders of magnitude lower than in typical antiferromagnetic materials \cite{Johnson1959, Hisamoto1960, Tinkham1963}. 

\begin{figure*}[!tbp]
\includegraphics[width=18cm]
{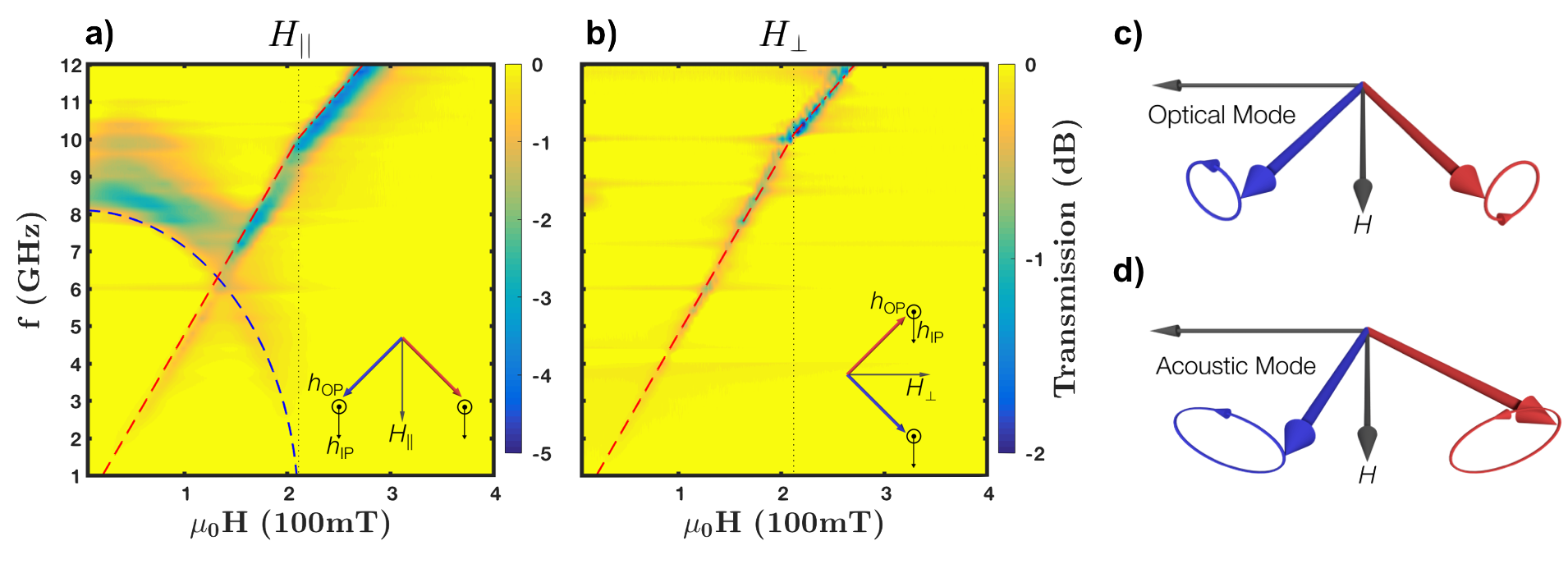}
\caption{ {\bf a)} and {\bf b)} Microwave transmission as a function of frequency and in-plane magnetic field at 1.56 K with, {\bf a)}, the magnetic field $H_{||}$ applied parallel to the in-plane RF field and, {\bf b)}, the magnetic field $H_{\perp}$ applied perpendicular to the in-plane RF field (the data for the two panels were taken from different samples). The regions of lower transmission arise from magnetic resonance. Two modes are observed in the $H_{||}$ configuration: an optical mode that has finite frequency at zero applied field, and an acoustic mode with frequency proportional to the applied field. Only the acoustic mode is observed in the $H_{\perp}$ configuration. The blue and red dashed lines in both panels are fits of the optical and acoustic mode frequencies to Eq.\ \ref{OpticalDispersion}  and Eq.\ \ref{AcousticDispersion} respectively. The insets of panels {\bf a)} and {\bf b)} show the relative orientation of the DC magnetic field, the equilibrium sublattice magnetizations, and the in-plane ($h_{\mathrm{IP}}$) and out-of-plane ($h_{\mathrm{OP}}$) components of the RF field. {\bf c)} and {\bf d)} Schematic illustrations of the precession orbits for the two sublattice magnetizations in, {\bf c)}, the optical mode and, {\bf d)}, the acoustic mode.}
\label{fig2}
\end{figure*}

Magnetic resonance measurements of CrCl$_3$ have a long history, including one of the earliest observations of paramagnetic resonance in a crystal \cite{Ramsey1999}. However, the dynamics below the N\'{e}el temperature remain largely unexplored. To study these dynamics, we first synthesized bulk CrCl$_3$ crystals according to the method of McGuire {\it et al.\ }\cite{McGuire22017, SI}.  The CrCl$_3$ platelets are transferred to a coplanar waveguide (CPW) and secured with polyamide (Kapton) tape (Fig. \ref{fig1}c). The crystal $c$ axis is normal to the CPW plane. The CPW is then mounted in a cryostat and connected to a Vector Network Analyzer by RF cables for microwave transmission measurements. A DC magnetic field is applied with the field directions illustrated in Fig. \ref{fig1}c. To study the response in the linear regime and prevent heating effects, we use a low power excitation signal (estimated to be -35 dBm at the sample).

We measure magnetic resonance by fixing the excitation frequency and sweeping the applied magnetic field. We observe distinct resonant features with different field geometries (Fig. \ref{fig1}d).  Only one resonant feature is observed when the DC magnetic field is applied perpendicular to the RF field ($H_{\perp}$), but two split features show up when the DC magnetic field is applied parallel to the RF field ($H_{||}$). To better understand this difference, we plot the transmission as a function of both excitation frequency and applied magnetic field (Fig. \ref{fig2}a, \ref{fig2}b). Under $H_{||}$, two modes exist with distinct field dependencies (Fig. \ref{fig2}a): one starting from finite frequency and softening with applied field, and the other with frequency proportional to the applied field. Remarkably, the modes cross without apparent interaction leading to a degeneracy at their crossing point; as we discuss below this crossing is protected by symmetry when the applied field lies in the crystal planes. With $H_{\perp}$, we see only the linearly dispersing mode (Fig. \ref{fig2}b). 

\begin{figure*}[!tbph]
\includegraphics[width=18cm]{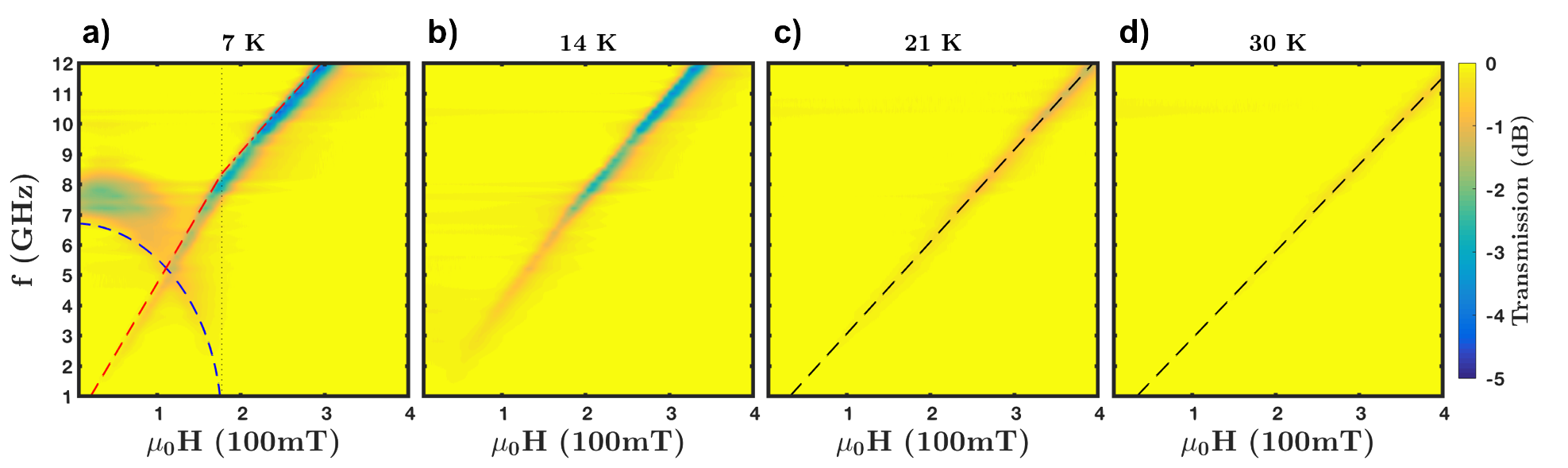}
    \caption{Microwave transmission as a function of frequency and in-plane magnetic field at 7 K, 14 K, 21 K, and 30 K. At 7 K, the sample is in antiferromagnetic state and both acoustic and optical modes are observed; $\mu_0 H_E=89$ mT and $\mu_0 M_s=323$ mT are determined by fits to Eqs. \ref{OpticalDispersion} and \ref{AcousticDispersion}, which are smaller than those at 1.56 K (Fig. \ref{fig2}a). Only one mode is observed at 14 K, and its frequency does not show a purely linear field dependence. At 21 K and 30 K, a single mode with linear field dependence is observed, arising from electron paramagnetic resonance. }
    \label{fig3}
\end{figure*}

To understand the origin of the two modes and the dependence on field geometry, we model the magnetic dynamics of CrCl$_3$ in the macrospin approximation. We assume that the magnetization direction is uniform within each layer, and introduce unit vectors $\hat{m}_{\mathrm{A}}$ and $\hat{m}_{\mathrm{B}}$ to represent the instantaneous direction of the magnetization on the $A$ and $B$ sublattices respectively. To account for the easy plane anisotropy of CrCl$_3$ \cite{Narath21965,McGuire22017}, we assume an out-of-plane demagnetization field of $\mu_0 M_{\mathrm{s}}$ with $M_{\mathrm{s}}$ being the effective saturation magnetization. The interlayer exchange energy is approximated as $\mu_0 M_{\mathrm{s}} H_{\mathrm{E}}\hat{m}_{\mathrm{A}} \cdot \hat{m}_{\mathrm{B}}$, where $H_{\mathrm{E}}$ is the interlayer exchange field. The energy depends only very weakly on the in-plane orientation of the magnetic moments \cite{Narath21965,McGuire22017} and we neglect the small in-plane anisotropy. Omitting the damping terms, we get a coupled Landau-Lifshitz-Gilbert (LLG) equation \cite{Keffer1952}:
\begin{align}
\small
\begin{split}
\frac{\mathrm{d}\hat{m}_{\mathrm{A}}}{\mathrm{d}t} &=-\mu_0\gamma\hat{m}_{\mathrm{A}}\times\left(\mathbf{H} -H_{\mathrm{E}}\hat{m}_{\mathrm{B}}-M_{\mathrm{s}}(\hat{m}_{\mathrm{A}}\cdot \hat{z})\hat{z}\right)+\boldsymbol{\tau}_{\mathrm{A}}\\
\frac{\mathrm{d}\hat{m}_{\mathrm{B}}}{\mathrm{d}t} &=-\mu_0\gamma\hat{m}_{\mathrm{B}}\times\left(\mathbf{H} -H_{\mathrm{E}}\hat{m}_{\mathrm{A}}- M_{\mathrm{s}}(\hat{m}_{\mathrm{B}}\cdot \hat{z})\hat{z}\right)+\boldsymbol{\tau}_{\mathrm{B}}.\label{FullLLG}
\end{split}
\end{align}
\normalsize
Here $\gamma$ is the gyromagnetic ratio, $\hat{z}$ is the direction perpendicular to sample plane (along the crystal $c$ axis). $\boldsymbol{\tau}_{\mathrm{A}}$ and $\boldsymbol{\tau}_{\mathrm{B}}$ are the torques which arise from the RF field of the CPW.

When the magnetic field, $\mathbf{H}$, is applied in the layer plane, Eq.\ \ref{FullLLG} is symmetric under twofold rotation around the applied field direction combined with sublattice exchange \cite{SI}. In the linear approximation, this results in two independent modes with even and odd parity under the symmetry (optical mode and acoustic mode, respectively, see Fig. \ref{fig2}c and \ref{fig2}d). The optical and acoustic modes result in Lorentzian resonances centered around the frequencies $\omega_{\pm}$. The frequencies have magnetic field dependence \cite{SI, Mandel1972, Streit1980}:
\begin{equation}
\omega_{+}=\mu_0\gamma\sqrt{2 H_{\mathrm{E}}M_{\mathrm{s}}\left(1-\frac{H^2}{4H^2_{\mathrm{E}}}\right)},\label{OpticalDispersion}
\end{equation}
and
\begin{equation}
\omega_{-}=\mu_0\gamma\sqrt{2H_{\mathrm{E}}(2H_{\mathrm{E}}+M_{\mathrm{s}})}\frac{H}{2H_{\mathrm{E}}}.\label{AcousticDispersion}
\end{equation}

Using Eqs. 2 and 3, we can fit the resonance frequencies of the different branches; these fits are shown by the dashed lines in Fig. \ref{fig2}a.  We find best fit values of $\mu_0 H_{\mathrm{E}}=105$ mT and $\mu_0M_{\mathrm{s}}=396$ mT at $T$ = 1.56 K, assuming $\gamma/2\pi \approx 28$ GHz/T for CrCl$_3$ \cite{Chehab1991}. The observed saturation magnetization is very close to 3$\mu_{\mathrm{B}}$ per Cr atom, consistent with previous magnetometry \cite{McGuire22017} and confirming that the out-of-plane crystalline anisotropy is negligible in this material. We also note that the acoustic branch changes its slope at $\mu_0 H\approx 200$ mT. This occurs because the moments of the two sublattices are aligned with the applied field direction when $H>2H_E$ \cite{SI}. In this case the crystal behaves as a ferromagnet and the acoustic mode transforms into uniform ferromagnetic resonance (FMR) with a resonant frequency described by the Kittel formula $\omega_{\mathrm{FMR}}=\mu_0\gamma \sqrt{H(H+M_s)}$. Figures \ref{fig2}a and \ref{fig2}b also show fits of the data for $H>2H_E$ to the Kittel formula (dash-dotted line). (The data above and below $H=2H_{\mathrm{E}}$ are fit simultaneously to extract a consistent parameter set.)

The dependence on field geometry in Fig.\ \ref{fig2} can now be understood as a consequence of selection rules for the even and odd parity modes. We can state the rule as follows: an RF magnetic field will excite the even (odd) parity mode if it is even (odd) under twofold rotation around the applied field direction \cite{SI}. The RF magnetic field generated from the CPW (Fig. \ref{fig1}c) has both in-plane and out-of-plane components. Directly over the signal line, the RF field points in the sample plane, while in the gap between the signal line and ground, the RF field is perpendicular to the sample plane. Our crystal is large enough to cover both regions and experience both field directions.  In the perpendicular geometry ($H_{\perp}$), both the in-plane and out-of-plane RF fields change sign under twofold rotation around the applied field direction (Inset of Fig. \ref{fig2}b). Therefore only the odd parity (acoustic) mode will be excited, as we observe. In the parallel field geometry ($H_{||}$), the in-plane component is invariant under the twofold rotation and excites the even parity (optical) mode, while the out-of-plane component changes sign and excites the odd-parity (acoustic) mode (Inset of Fig. \ref{fig2}a). We will focus on measurements in the parallel field geometry because it allows simultaneous excitation of both modes. 

\begin{figure*}[!tbph]
\includegraphics[width=\textwidth]{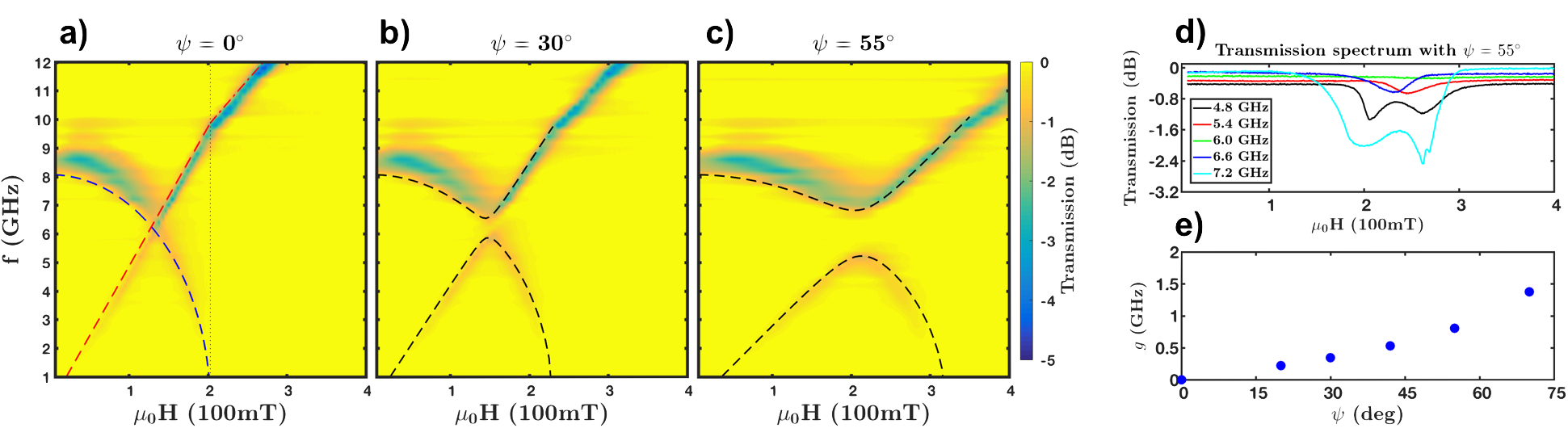}
    \caption{ {\bf a) b) c)} Microwave transmission as a function of frequency and applied field; the field is applied at an angle of $\psi=0^{\circ}$, $30^{\circ}$, and $55^{\circ}$, from the sample plane in panel {\bf a)}, {\bf b)}, and {\bf c)}. When the field is applied in-plane, the mode crossing is protected by rotational symmetry combined with sublattice exchange. Out-of-plane field breaks the symmetry and couples the two modes resulting in large tunable coupling gaps. {\bf d)} Microwave transmission versus applied field at $\psi=55^{\circ}$ for various frequencies, showing the coupling gap. {\bf e)} The coupling strength $g$ increases with $\psi$, and can be tuned from 0-1.37 GHz. }
    \label{fig4}
\end{figure*}

We further study the evolution of the AFMR signal as a function of temperature. As the temperature is increased from 1.56K (Fig. \ref{fig2}a) to 7K (Fig. \ref{fig3}a), the optical mode frequency decreases due to the reduction of $H_{\mathrm{E}}$ and $M_{\mathrm{s}}$. The optical mode disappears entirely at 14 K, implying that the sample is no longer antiferromagnetic, consistent with previous measurements of the N{\'e}el temperature \cite{McGuire22017}. At higher temperatures, the magnetic resonance frequency depends linearly on the applied field with a slope of 30.4 GHz/T and 28.8 GHz/T at 21 K and 30 K, respectively (Fig. \ref{fig3}c, d). This is electron paramagnetic resonance arising from Cr$^{3+}$ ions that has been reported previously \cite{Chehab1991}. Unlike a conventional antiferromagnet, the Curie-Weiss temperature for CrCl$_3$ is positive due to the large ferromagnetic intralayer exchange \cite{McGuire22017}; this leads to a very large magnetic susceptibility just above N{\'e}el temperature. In this temperature range, the magnetization is proportional to the applied field through $M=\chi\left(T\right) H$. Then the resonant frequency is $\omega=\mu_0\gamma \sqrt{H\left(H+M\left(H, T\right)\right)}=\mu_0\gamma\sqrt{1+\chi\left(T\right)}H$ \cite{stanger_role_1997}. Our data correspond to $\chi(21\,\mathrm{K})=0.178$ and $\chi(30\,\mathrm{K})=0.056$. For the temperature range 14-20 K, the resonant frequencies do not have a purely linear dependence on applied field (Fig. \ref{fig3}b). This is likely due to a non-linear relationship $M(H)$ as $M$ approaches its saturation value, previously detected in magnetization and magneto-optical experiments \cite{McGuire22017,Kuhlow1982}.

So far, we have discussed AFMR with an in-plane applied magnetic field \cite{Mandel1972}. In this case, the system is symmetric under twofold rotation around the applied field direction combined with sublattice exchange. This symmetry prevents hybridization between the optical and acoustic modes and leads to a degeneracy where they cross, as seen in Fig. \ref{fig2}a and \ref{fig3}a. In principle, breaking this symmetry can hybridize the two modes and generate an anti-crossing gap. As suggested by Eq.\ \ref{FullLLG}, one possible approach for inducing such a symmetry breaking is to utilize different $M_{\mathrm{s}}$ for the $A$ and $B$ sublattices by stacking different 2D magnets. Here, we instead employ an out-of-plane field to break the $180^{\circ}$ rotational symmetry. To test this latter concept, we measure the AFMR spectrum for a DC magnetic field applied at a range of angles, $\psi$, from the CPW plane. For $\psi = 30^{\circ}$, we see that optical and acoustic mode structure is largely unchanged, except that a gap opens near the crossing point (Fig. \ref{fig4}b). Increasing the tilt angle increases the gap size as shown in Fig. \ref{fig4}c. Therefore, by breaking the rotational symmetry with an out-of-plane field, we can introduce a magnon-magnon coupling between the previously uncoupled magnon modes. 

To provide a quantitative description of this magnon-magnon coupling, we turn to the matrix formalism of the LLG equation \cite{SI}. The result is high and low frequency branches of antiferromagnetic resonance, continuously connected to the even and odd parity modes. The evolution of both modes and their mixing can be captured by the eigenvalue problem of a two-by-two matrix:
\begin{equation}
\begin{vmatrix}
\omega^2_{\mathrm{a}}(H,\psi)-\omega^2 & \Delta^2(H,\psi) \\
\Delta^2(H,\psi) & \omega^2_{\mathrm{o}}(H,\psi)-\omega^2 
\end{vmatrix}=0 \label{GapEquation}
\end{equation}
Here $\omega_{\mathrm{a}}=\mu_0\gamma\sqrt{1+\frac{M_{\mathrm{s}}}{2H_{\mathrm{E}}}}H\cos\psi$ is the bare acoustic mode frequency and  $\omega_{\mathrm{o}}=\mu_0\gamma\sqrt{2H_{\mathrm{E}}M_{\mathrm{s}}\left(1-\frac{H^2}{H^2_{\mathrm{FM}}}\right)+\frac{\sin^2\psi}{\left(1+\frac{M_{\mathrm{s}}}{2H_{\mathrm{E}}}\right)^2}H^2}$ is the bare optical mode frequency. $H_{\mathrm{FM}}$ is the applied magnetic field required to fully align the two sublattices, satisfying $1/H^2_{\mathrm{FM}}=\cos^2\psi/(2H_{\mathrm{E}})^2+\sin^2\psi/(2H_{\mathrm{E}}+M_{\mathrm{s}})^2$. $\Delta$ represents the magnon-magnon coupling term which turns the accidental degeneracy of the two modes into an avoided crossing. $\Delta=\mu_0\gamma H\left(\frac{2H_{\mathrm{E}}}{2H_{\mathrm{E}}+M_{\mathrm{s}}}\sin^2\psi \cos^2\psi\right)^{1/4}$. The solutions, $\omega$, of Eq.\ \ref{GapEquation} are the resonance frequencies of the LLG equation. When $|\omega_{\mathrm{o}}-\omega_{\mathrm{a}}|\gg \Delta $, the effect of the coupling term is negligible and the mode frequencies are approximately $\omega\approx\omega_{\mathrm{o}}$ and  $\omega\approx\omega_{\mathrm{a}}$. When the optical and acoustic modes become closer in frequency, they are hybridized by the coupling term opening a gap. This coupling is zero for $\psi=0$ and only becomes non-zero as we cant the applied field out-of-plane. $\Delta=\mu_0\gamma H\left(\frac{2H_{\mathrm{E}}}{2H_{\mathrm{E}}+M_{\mathrm{s}}}\sin^2\psi \cos^2\psi\right)^{1/4}$

The dashed lines in Fig. \ref{fig4}a, \ref{fig4}b, and \ref{fig4}c indicate fits to the eigenvalues of Eq.\ \ref{GapEquation}. The coupling strength of the two modes, $g$, is determined as half of the minimal frequency spacing in the fits. We can also extract the dissipation rates of the upper and lower branches, $\kappa_{\mathrm{U}}$ and $\kappa_{\mathrm{L}}$, by Lorentzian fitting of the frequency dependence of the transmission. For $\psi=55^{\circ}$, we obtain $g=0.8$ GHz, $\kappa_{\mathrm{U}}\approx 0.69$ GHz, and $\kappa_{\mathrm{L}}\approx 0.15$ GHz which indicates that strong magnon-magnon coupling is achieved as $g > \kappa_{\mathrm{U}}$ and $g > \kappa_{\mathrm{L}}$ \cite{chen_strong_2018}. The cooperativity is $C=g^2/(\kappa_{\mathrm{U}} \times \kappa_{\mathrm{L}})=6.2$, which is large and can be improved by using a more homogeneous sample. Fig. \ref{fig4}e shows the angular dependence of $g$, which monotonically increases with $\psi$. By simply rotating the crystal alignment in an applied field, a tunable coupling is realized that can tune the system from a symmetry-protected mode crossing to the strong coupling regime. 

In summary, we have measured magnetic resonance of the layered antiferromagnet CrCl$_3$ as a function of temperature and applied magnetic field, with the magnetic field applied at a variety of angles from the crystal planes. We have shown that CrCl$_3$ possess an unusually rich GHz-frequency AFMR spectrum due to the weak interlayer coupling. We detect both acoustic and optical branches of AFMR and show that an applied magnetic field can induce an accidental degeneracy between them. Furthermore, by breaking rotational symmetry we can induce a coupling between these modes and open a tunable gap. All of these effects are captured with analytical solutions to the LLG equation. While we have focused on the small-angle dynamics here, we expect interaction between the modes in the nonlinear regime. For example, three-magnon processes could potentially be triggered when the frequency of acoustic and optical modes satisfy certain relationships. 

There is also tremendous interest in using mechanical exfoliation to isolate ultrathin layered magnets down to the monolayer limit \cite{Huang2017, Gong2017}, and to incorporate them in van der Waals heterostructures \cite{Klein2018, Song2018}. Because CrCl$_3$ can be cleaved to produce air-stable and atomically thin films \cite{McGuire22017}, we expect our results to enable a new generation of device-based antiferromagnetic spintronics with microwave control of the N{\'e}el vector. Beyond CrCl$_3$, our results apply broadly within the class of transition metal trihalides, so that the frequency scale can be tuned by varying the chemical composition and thickness \cite{McGuire22017}. Using van der Waals assembly, we can combine different magnetic materials to access further tunability and even induce magnon-magnon coupling at zero applied field by breaking sublattice exchange symmetry.

This work was supported by the Center for Integrated Quantum Materials under NSF Grant DMR-1231319 (D.R.K.), the DOE Office of Science, Basic Energy Sciences under award $\text{DE-SC0018935}$ (D.M.), as well as the Gordon and Betty Moore Foundation'€™s EPiQS Initiative through grant GBMF4541 to P.J.-H. D.R.K. acknowledges partial support by the NSF Graduate Research Fellowship Program under Grant No. 1122374. J.T.H., P.Z., and L.L. acknowledge support from National Science Foundation under award ECCS-1808826.

\bibliography{CrCl3_AFMR}

\newpage
\onecolumngrid

\section*{Supplemental Material}

\centerline{\bf Sample Preparation and Data Acquisition}

We load approximately 1 g of 99.9\% purity anhydrous CrCl$_3$ flakes (Alpha Aesar) into a silica ampoule (16 mm inner diameter, 55 cm in length) in an argon environment, followed by evacuation and sealing of the ampoule. The ampoule is placed in a three-zone horizontal tube furnace. For the growth period, the source zone is heated to 700$^{\circ}$C, the middle zone is heated to 550$^{\circ}$C and the final zone is heated to 625$^{\circ}$C. These temperatures are maintained for 6 days, after which we find that the entire source material is recrystallized as large platelets in the middle of the tube. The CrCl$_3$ platelets are then transferred to a coplanar waveguide (CPW) and secured with polyamide (Kapton) tape for microwave transmission measurements. The signal line in CPW is 0.94 mm wide, and the gap between signal line and ground is 0.15 mm wide. The 2D plots in this paper are obtained by combining transmission as a function of magnetic field at various frequencies. For each fixed frequency trace, we subtract the transmission average over the range from 440-450 mT, to remove the frequency dependent background signal. \\

\centerline{\bf Symmetry and Solutions of the LLG Equation}

\begin{figure}[!tbph]
\includegraphics[width=8.5cm]{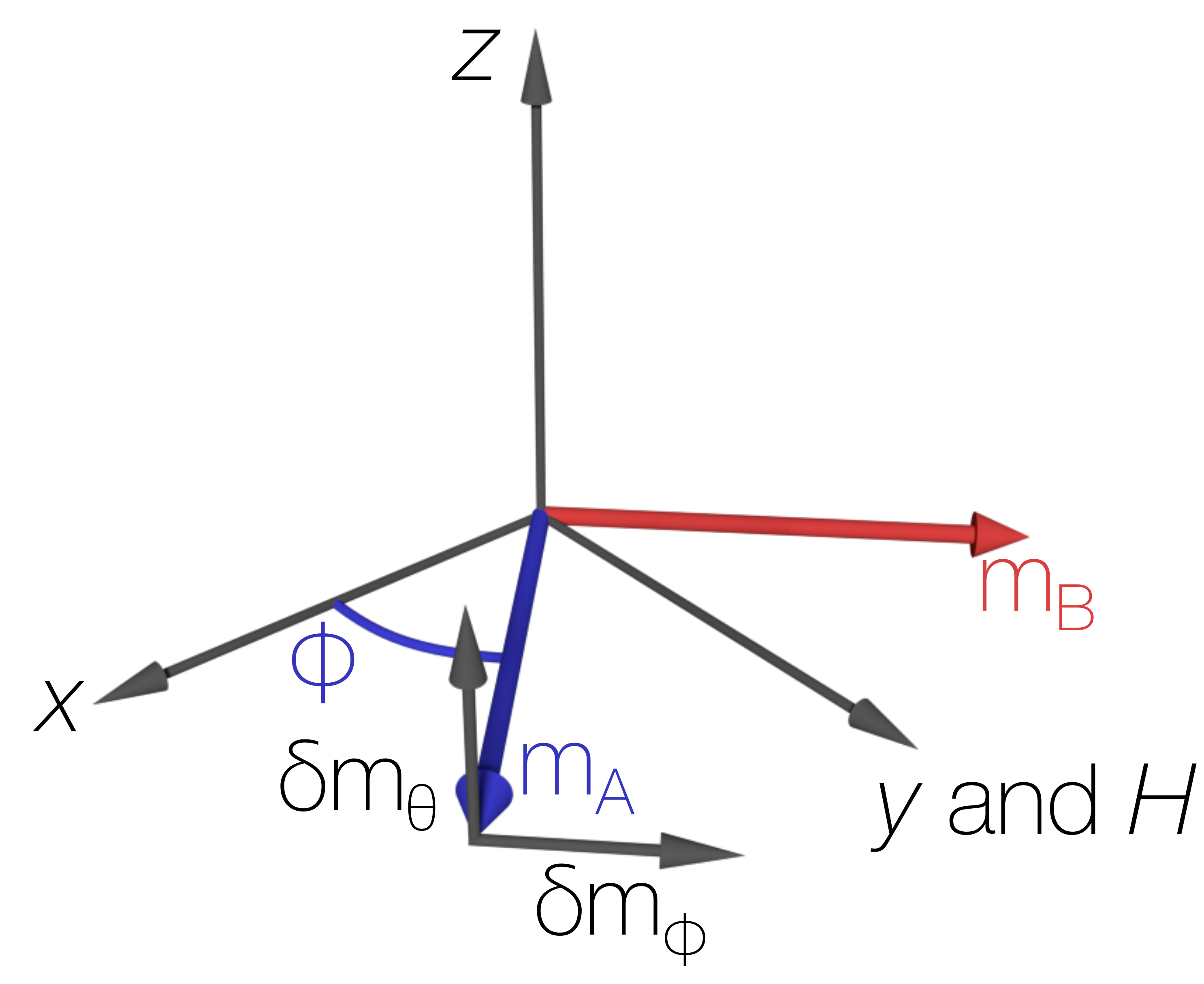}
    \caption{ Illustration of the basis used for solving the LLG equation when the DC magnetic field is applied in the $xy$ plane. The field is applied along the $y$ direction. Above the spin-flop field ($\approx 20$ mT) the N\'{e}el vector will orient perpendicular to the external field and the moments will cant towards it at an angle $\phi$.} \label{Supp_fig_0}
\end{figure}

Our first goal here is to solve Eq.\ \ref{FullLLG} in the approximation of small precession angle. For the magnetic field applied in the sample plane, we will use the coordinates defined in Fig. \ref{Supp_fig_0}. We expand $\hat{m}_{\mathrm{A}}=\hat{m}^{\mathrm{eq}}_{\mathrm{A}}+\delta\mathbf{m}_{\mathrm{A}}\exp{i\omega t}$ and likewise for the $B$ sublattice magnetization. We assume harmonic time dependence at frequency $f=\omega/2\pi$ of the torques. In equilibrium, the moments orient perpendicular to the applied field in a spin-flop transition; even a small external field ($\approx20$ mT) suffices to effect this reorientation \cite{McGuire22017}. We assume $\hat{m}^{\mathrm{eq}}_{\mathrm{A}}=\hat{x}$ at zero applied field. As the applied field is increased $\hat{m}^{\mathrm{eq}}_{\mathrm{A}}$ and $\hat{m}^{\mathrm{eq}}_{\mathrm{B}}$ will cant at an angle $\phi$ towards it. This angle can be shown to satisfy $\sin\phi=H/2H_{\mathrm{E}}$. Note also that the problem is symmetric under twofold rotational symmetry around the $y$ axis combined with sublattice exchange, implying that $C_{2y}\hat{m}^{\mathrm{eq}}_{\mathrm{A}}=\hat{m}^{\mathrm{eq}}_{\mathrm{B}}$, where $C_{2y}$ rotates vectors by 180$^{\circ}$ around the $y$ axis. We will discuss the properties of this symmetry in more detail later. 

Substituting in Eq.\ \ref{FullLLG} and keeping linear-order terms gives:
\begin{align}
\begin{split}
i\omega\delta\mathbf{m}_{\mathrm{A}} &=\mu_0\gamma\hat{m}^{\mathrm{eq}}_{\mathrm{A}} \times \left(H_{\mathrm{eq}}\delta\mathbf{m}_{\mathrm{A}}+H_{\mathrm{E}}\delta\mathbf{m}_{\mathrm{B}}+M_{\mathrm{s}}(\delta\mathbf{m}_{\mathrm{A}}\cdot \hat{z})\hat{z} \right) +\boldsymbol{\tau}_{\mathrm{A}} \\
i\omega\delta\mathbf{m}_{\mathrm{B}}&= \mu_0\gamma \hat{m}^{\mathrm{eq}}_{\mathrm{B}} \times\left(H_{\mathrm{eq}}\delta\mathbf{m}_{\mathrm{B}}+H_{\mathrm{E}}\delta\mathbf{m}_{\mathrm{A}}+M_{\mathrm{s}}(\delta\mathbf{m}_{\mathrm{B}}\cdot \hat{z})\hat{z} \right)+\boldsymbol{\tau}_{\mathrm{B}}.\label{LinearLLG}
\end{split}
\end{align}
Here $H_{\mathrm{eq}}$ is the positive number such that $H\hat{y}-H_{\mathrm{E}}\hat{m}^{\mathrm{eq}}_{\mathrm{B}}=H_{\mathrm{eq}}\hat{m}^{\mathrm{eq}}_{\mathrm{A}}$. (This number exists because the left-hand-side points along $\hat{m}_{\mathrm{A}}$ in equilibrium.) Therefore, $H_{\mathrm{eq}}=|H\hat{y}-H_{\mathrm{E}}\hat{m}^{\mathrm{eq}}_{\mathrm{B}}|=H_{\mathrm{E}}$ independent of the applied field. The final equality uses $\hat{y}\cdot\hat{m}^{\mathrm{eq}}_{\mathrm{B}}=\sin\phi=H/2H_{\mathrm{E}}$.

To decouple the equations we act $C_{2y}$ on both sides of the second line of Eq.\ \ref{LinearLLG} and add it to the first. This creates an independent equation for $\delta \mathbf{m}_{+}=\delta\mathbf{m}_{\mathrm{A}}+C_{2y}\delta\mathbf{m}_{\mathrm{B}}$ which couples to a linear combination of the torques $\boldsymbol{\tau}_{+}=\boldsymbol{\tau}_{\mathrm{A}}+ C_{2y}\boldsymbol{\tau}_{\mathrm{B}}$. Likewise we can form an equation for $\delta \mathbf{m}_{-}=\delta\mathbf{m}_{\mathrm{A}}-C_{2y}\delta\mathbf{m}_{\mathrm{B}}$ excited by $\boldsymbol{\tau}_{-}=\boldsymbol{\tau}_{\mathrm{A}}- C_{2y}\boldsymbol{\tau}_{\mathrm{B}}$. The action of $C_{2y}$ is simplified by noting that:
\begin{align}
\begin{split}
 & C_{2y}  \left[\hat{m}^{\mathrm{eq}}_{\mathrm{B}}  \times  \left(H_{\mathrm{E}}\delta\mathbf{m}_{\mathrm{B}}+H_{\mathrm{E}}\delta\mathbf{m}_{\mathrm{A}}+M_{\mathrm{s}}\left(\delta\mathbf{m}_{\mathrm{B}}\cdot \hat{z}\right)\hat{z} \right)\right] = \\
& \left(C_{2y} \hat{m}^{\mathrm{eq}}_{\mathrm{B}}\right)  \times  \left(H_{\mathrm{E}}C_{2y}\delta\mathbf{m}_{\mathrm{B}}+H_{\mathrm{E}}C_{2y}\delta\mathbf{m}_{\mathrm{A}}-M_{\mathrm{s}}\left(\delta\mathbf{m}_{\mathrm{B}}\cdot \hat{z}\right)\hat{z} \right)
\end{split}
\end{align}
i.e. that the action of $C_{2y}$ on a cross-product of two vectors is equivalent to rotating the two vectors and then taking their cross-product. Using such identities we derive the decoupled equations for the even and odd parity modes:
\begin{align}
\begin{split}
i\omega\delta\mathbf{m}_{\pm} &= \\
 \mu_0\gamma \hat{m}^{\mathrm{eq}}_{\mathrm{A}} & \times \left(H_{\mathrm{E}}\delta\mathbf{m}_{\pm}\pm H_{\mathrm{E}}C_{2y}\delta\mathbf{m}_{\pm}+M_{\mathrm{s}}(\delta\mathbf{m}_{\pm}\cdot \hat{z})\hat{z} \right) +\boldsymbol{\tau}_{\pm}.\label{FinalLLG}
\end{split}
\end{align}
To include magnetic damping, we can add a term $i\omega\alpha\hat{m}^{\mathrm{eq}}_{\mathrm{A}}\times\delta\mathbf{m}_{\pm}$. Equation \ref{FinalLLG} is solved using the standard methods relevant for a ferromagnet in the macrospin approximation \cite{Kittel1948}. The resonant frequencies obtained this way are given in Eqs. \ref{OpticalDispersion} and \ref{AcousticDispersion} of the main text.

Given a solution of Eq.\ \ref{FinalLLG}, we can recover the motions of the sublattice moments using $\delta \mathbf{m}_{\mathrm{A}}=(\delta\mathbf{m}_{+}+\delta\mathbf{m}_{-})/2$ and $\delta\mathbf{m}_{\mathrm{B}}=C_{2y}(\delta\mathbf{m}_{+}-\delta\mathbf{m}_{-})/2$. Looking at pure excitation of the $+$ mode, we have $\delta\mathbf{m}_{\mathrm{A}}=C_{2y}\delta\mathbf{m}_{\mathrm{B}}$ (see Fig. \ref{fig2}c). On the other hand, pure excitation of the $-$ mode gives $\delta\mathbf{m}_{\mathrm{A}}=-C_{2y}\delta\mathbf{m}_{\mathrm{B}}$. We can also figure out the selection rules for excitation of the two modes. For pure $\delta\mathbf{m}_{\mathrm{+}}$ excitation $\boldsymbol{\tau}_{-}=0$ so that $\boldsymbol{\tau}_{\mathrm{A}}=C_{2y}\boldsymbol{\tau}_{\mathrm{B}}$. In terms of the RF field $\boldsymbol{\tau}_{\mathrm{B}}=-\mu_0\gamma M_{\mathrm{s}}\hat{m}^{\mathrm{eq}}_{\mathrm{B}} \times \mathbf{h}_{\mathrm{RF}}$ and $C_{2y}\boldsymbol{\tau}_{\mathrm{B}}=-\mu_0\gamma M_{\mathrm{s}}\hat{m}^{\mathrm{eq}}_{\mathrm{A}} \times C_{2y}\mathbf{h}_{\mathrm{RF}}$. Thus we recover the condition stated in the main text that pure excitation of the $+$ mode requires $\mathbf{h}_{\mathrm{RF}}=C_{2y}\mathbf{h}_{\mathrm{RF}}$. We can similarly show that excitation of the $-$ mode requires $\mathbf{h}_{\mathrm{RF}}=-C_{2y}\mathbf{h}_{\mathrm{RF}}$. 

Before solving the LLG equation in the general case where the magnetic field is applied in an arbitrary direction, we discuss the symmetry properties of Eq.\ \ref{LinearLLG} in more detail. The equation can be re-written as a $4\times 4$ matrix in the basis $\left(\delta m_{+,\phi}, \delta m_{+,\theta},\delta m_{-,\phi}, \delta m_{-,\theta} \right)$ (see Fig. \ref{Supp_fig_0} for the definition of this basis). In this basis it is block diagonal:
\begin{equation}
i\omega\begin{pmatrix}
\delta\mathbf{m}_{+}\\
\delta\mathbf{m}_{-}
\end{pmatrix}
=\begin{pmatrix}
A & 0 \\
0 & B
\end{pmatrix}
\begin{pmatrix}
\delta\mathbf{m}_{+}\\
\delta\mathbf{m}_{-}
\end{pmatrix},\label{LLGMatrixIP}
\end{equation}
where
\begin{equation}
A=\begin{pmatrix}
0 & -M_{\mathrm{s}} \\
2H_{\mathrm{E}}\cos^2\phi & 0
\end{pmatrix},
\end{equation}
and
\begin{equation}
B=\begin{pmatrix}
0 & -2H_{\mathrm{E}}-M_{\mathrm{s}} \\
2H_{\mathrm{E}}\sin^2\phi & 0
\end{pmatrix}.
\end{equation}
The LLG matrix commutes with the following $4\times 4$ matrix:
\begin{equation}
\Sigma=\begin{pmatrix}
1 & 0 & 0 & 0 \\
0 & 1 & 0 & 0 \\
0 & 0 & -1 & 0 \\
0 & 0 & 0 & -1
\end{pmatrix}.
\end{equation}
This turns out to be just the combination of twofold rotation and sublattice exchange that we have previously discussed. To see this consider a state with displacements $\delta \mathbf{m}_{\mathrm{A}}$ and $\delta \mathbf{m}_{\mathrm{B}}$. Now consider a new state with the displacement $C_{2y}\delta \mathbf{m}_{\mathrm{A}}$ on the $B$ sublattice and displacement $C_{2y}\delta \mathbf{m}_{\mathrm{B}}$ on the $A$ sublattice (i.e. act on the system with twofold rotation and sublattice exchange). We can calculate $\delta \mathbf{m}_{+}$ before and after the transformation. Before it is $\delta \mathbf{m}_{\mathrm{A}}+C_{2y}\delta \mathbf{m}_{\mathrm{B}}$ and afterwards $C_{2y}\delta \mathbf{m}_{\mathrm{B}}+C_{2y}C_{2y}\delta \mathbf{m}_{\mathrm{A}}$; therefore, $\delta \mathbf{m}_{+}$ is invariant under the combination of twofold rotation and sublattice exchange. Similarly $\delta \mathbf{m}_{-}$ is odd under the transformation. This is just the relation described by $\Sigma$ above. In any case, we can choose the eigenmodes of the linearized LLG equation to also be eigenvectors of $\Sigma$ with eigenvalue $\pm 1$. One can also show that adding any term to the LLG equation that commutes with $\Sigma$ will not mix modes with different $\Sigma$ eigenvalues. This leads to the symmetry-protected degeneracy that we have discussed in the main text and observed in CrCl$_3$. 

The above calculations apply to the case where the magnetic field is applied purely in-plane. When the magnetic field is applied at an angle $\psi$ from the $xy$ plane, the twofold rotation is broken and we can no-longer easily decouple the equations into independent modes. That requires us to solve the coupled equations with four degrees of freedom (rather than the simpler decoupled equations with two degrees of freedom each).

\begin{figure}[!tbph]
\includegraphics[width=8.5cm]{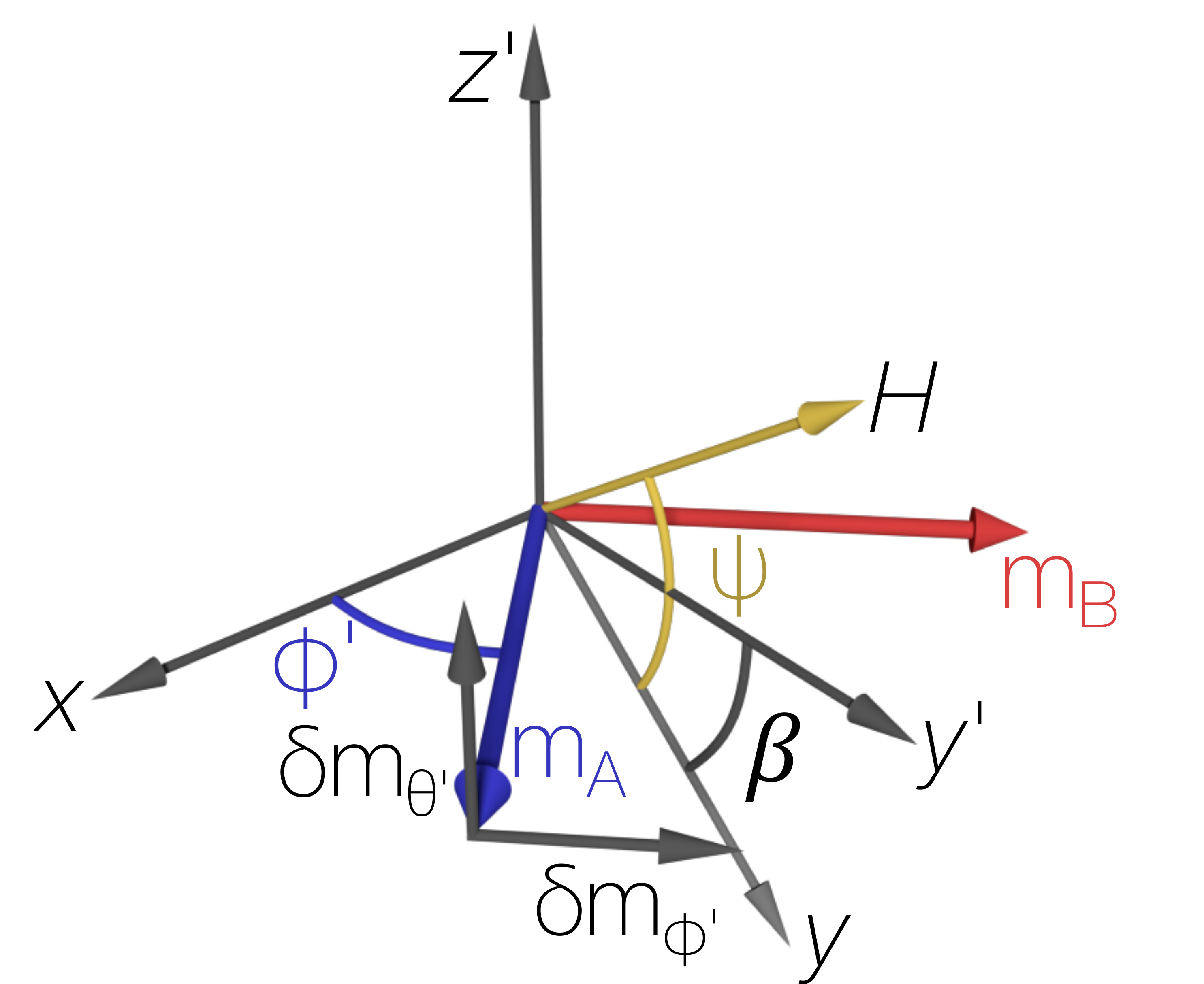}
    \caption{ Illustration of the basis used for solving the LLG equation when the DC magnetic field is applied at an angle of $\psi$ from the $xy$ plane. The new basis vector $\hat{y}'$ points along the direction $\hat{m}^{\mathrm{eq}}_{\mathrm{A}}+\hat{m}^{\mathrm{eq}}_{\mathrm{B}}$ which is not necessarily aligned with the applied field due to the $xy$ easy plane. } \label{Supp_fig}
\end{figure}

We find that the coupled equations take a relatively simple form in the basis illustrated in Fig. \ref{Supp_fig}. We set $\hat{y}'$ as the $H$ and $\psi$ dependent net magnetization direction, so that it points along $\hat{m}^{\mathrm{eq}}_{\mathrm{A}}+\hat{m}^{\mathrm{eq}}_{\mathrm{B}}$. The other basis vectors are $\hat{x}'=\hat{x}$ and $\hat{z}'=\hat{x}\times \hat{y}'$. Considering $C_{2y'}$ as a rotation around the $y'$ axis, then $\hat{m}^{\mathrm{eq}}_{\mathrm{A}}=C_{2y'}\hat{m}^{\mathrm{eq}}_{\mathrm{B}}$. Following the same steps required to derive Eq.\ \ref{FinalLLG}, we derive the equation for the dynamics of the even and odd parity modes (under $C_{2y'}$) in the general case:
\begin{align}
\begin{split}
i\omega\delta\mathbf{m}_{\pm} &= \mu_0\gamma \hat{m}^{\mathrm{eq}}_{\mathrm{A}}  \times \left(H_{\mathrm{eq}}\delta\mathbf{m}_{\pm}\pm H_{\mathrm{E}}C_{2y'}\delta\mathbf{m}_{\pm}\right)\\ 
& +\mu_0\gamma\hat{m}^{\mathrm{eq}}_{\mathrm{A}}  \times\left(\frac{M_{\mathrm{s}}}{2} \left(\hat{z} \hat{z}\cdot + C_{2y'}\hat{z} (C_{2y'}\hat{z})\cdot\right)\delta\mathbf{m}_{\pm}\right) \\
& +\mu_0\gamma\hat{m}^{\mathrm{eq}}_{\mathrm{A}}  \times\left(\frac{M_{\mathrm{s}}}{2} \left(\hat{z} \hat{z}\cdot - C_{2y'}\hat{z} (C_{2y'}\hat{z})\cdot\right)\delta\mathbf{m}_{\mp}\right).
\label{CoupledLLG}
\end{split}
\end{align}
We have also calculated that $H_{\mathrm{eq}}=H_{\mathrm{E}}$ even when the field is applied at an angle to the $xy$ plane, as long as the crystal is in the antiferromagnetic state, $H<H_{\mathrm{FM}}$. We introduce $\hat{z}\cdot$ and $(C_{2y'}\hat{z})\cdot$ as operators that act on $\delta\mathbf{m}_{\pm}$ by taking the dot product with $\hat{z}$ and $C_{2y'}\hat{z}$ respectively. Only the last cross-product couples the even and odd parity modes $\delta\mathbf{m}_{\pm}$; however the introduction of these modes is just a computational convenience and they are not weakly coupled over any range of applied magnetic fields when $0<\psi<\pi/2$. The actual weakly coupled modes will emerge below after further calculations. 

Equation \ref{CoupledLLG} defines a $4\times 4$ matrix mapping the vector of displacements $\left(\delta m_{+,\phi'}, \delta m_{+,\theta'},\delta m_{-,\phi'}, \delta m_{-,\theta'} \right)$ onto the vector of time derivatives, in the basis defined in Fig. \ref{Supp_fig}. The matrix equation is:

\begin{equation}
i\omega\begin{pmatrix}
\delta\mathbf{m}_{+}\\
\delta\mathbf{m}_{-}
\end{pmatrix}
=\begin{pmatrix}
A' & C \\
C & B'
\end{pmatrix}
\begin{pmatrix}
\delta\mathbf{m}_{+}\\
\delta\mathbf{m}_{-}
\end{pmatrix}.\label{LLGMatrix}
\end{equation}
$A'$, $B'$, and $C$ are $2\times 2$ matrices:
\begin{equation}
A'=\begin{pmatrix}
0 & -M_{\mathrm{s}}\cos^2\beta \\
2H_{\mathrm{E}}\cos^2\phi'+M_{\mathrm{s}}\sin^2\beta \cos^2\phi' & 0
\end{pmatrix},
\end{equation}
\begin{equation}
\begin{split}
B'=
\begin{pmatrix}
0 & -2H_{\mathrm{E}}-M_{\mathrm{s}}\cos^2\beta \\
2H_{\mathrm{E}}\sin^2\phi'+M_{\mathrm{s}}\sin^2\beta \cos^2\phi' & 0
\end{pmatrix},
\end{split}
\end{equation}
and
\begin{equation}
C=\begin{pmatrix}
-M_{\mathrm{s}}\cos\beta\sin\beta \cos\phi' & 0 \\
0 & M_{\mathrm{s}}\cos\beta\sin\beta \cos\phi'
\end{pmatrix}.
\end{equation}
Here $\phi'$ and $\beta$ are angles determined by the equilibrium sublattice magnetizations as defined in Fig. \ref{Supp_fig}. The various trigonometric functions of $\beta$ and $\phi'$ can be related to the angles in spherical coordinates $\hat{m}^{\mathrm{eq}}_{\mathrm{A}}=\cos\theta\cos\phi \hat{x}+\cos\theta\sin\phi \hat{y} +\sin\theta \hat{z}$. One can also show that the equilibrium coordinates satisfy $\sin\theta=H\sin\psi/(2H_{\mathrm{E}}+M_{\mathrm{s}})$ and $\sin\phi=H\cos\psi/(2H_{\mathrm{E}}\cos\theta)$.

We use computer algebra to calculate the determinant after subtracting $i\omega 1_4$, with $1_4$ being the identity matrix. The resulting equation for the eigenvalues is:
\begin{equation}
\begin{split}
\omega^4 -\left[\omega^2_{\mathrm{o}}(H,\psi)+\omega^2_{\mathrm{a}}(H,\psi)\right]\omega^2 
+\omega^2_{\mathrm{o}}(H,\psi)\omega^2_{\mathrm{a}}(H,\psi)-\Delta^4(H,\psi)=0,
\end{split}
\end{equation}
where the functions $\omega_{\mathrm{o}}$, $\omega_{\mathrm{a}}$ and $\Delta$ are defined in the main text. To gain intuition into the mode structure and to define the optical and acoustic branches, we simply note that this is the eigenvalue equation for the $2\times2$ matrix defined in Eq.\ \ref{GapEquation}. Therefore, the resonant dynamics can be described as two approximately independent resonances, with frequencies $\omega_{\mathrm{o}}$ and $\omega_{\mathrm{a}}$, coupled through a term $\Delta$ that becomes relevant when $\omega_{\mathrm{o}}$ and $\omega_{\mathrm{a}}$ become similar.

\end{document}